\documentclass[final]{dmtcs-episciences}

\usepackage[utf8]{inputenc}
\usepackage{subfigure}

\usepackage{amsmath}
\usepackage{amssymb}
\usepackage{subfiles}
\usepackage{color}
\usepackage{tikz}
\usetikzlibrary{shapes}
\usetikzlibrary{positioning}

\usepackage{amsthm}
\usepackage{thmtools}
\usepackage{mathtools}

\usepackage{breakurl}

\newtheorem{introtheorem}{Theorem}

\declaretheorem[numberwithin=section, style=plain, name=Theorem]{theorem}
\declaretheorem[sibling=theorem, style=plain, name=Lemma]{lemma}

\declaretheorem[sibling=theorem, style=definition, name=Definition]{definition}

\declaretheorem[sibling=theorem, style=remark, name=Example]{example}
\declaretheorem[numbered=no, style=remark, name=Remark]{remark}

\DeclareMathOperator{\sgon}{sgon}
\DeclareMathOperator{\gon}{gon}

\DeclareMathOperator{\tw}{tw}
\DeclareMathOperator{\poly}{poly}

\newcommand{\PP}{\ensuremath{\mathbf{P}}}

\newcommand{\case}[1]{\underline{#1:}}
\DeclareRobustCommand{\stirling}{\genfrac\{\}{0pt}{}}
\DeclarePairedDelimiter{\floor}{\lfloor}{\rfloor}
\DeclarePairedDelimiter{\parens}{(}{)}
\DeclarePairedDelimiter{\brackets}{[}{]}

\tikzstyle{vertex}=[circle, draw, fill=black, inner sep=0pt, minimum width=4pt]
\tikzstyle{added}=[diamond, fill=gray, inner sep=1.5pt, minimum width=4pt]
\tikzstyle{edge} = [line width = 1pt]

\author{Ragnar {Groot Koerkamp}\affiliationmark{1}
	\and Marieke {van der Wegen}\affiliationmark{1,2}}
\affiliation{
	Mathematical Institute, Utrecht University, The Netherlands\\
	Department of Information and Computing Sciences, Utrecht University, The Netherlands}

\title{Stable gonality is computable}


\received{2018-10-30}
\revised{2019-4-19}
\accepted{2019-4-22}

\begin{document}
	\publicationdetails{21}{2019}{1}{10}{4931}
	\maketitle
	\begin{abstract}
	\emph{Stable gonality} is a multigraph parameter that measures the complexity of a graph. It is defined using maps to trees. Those maps, in some sense, divide the edges equally over the edges of the tree; stable gonality asks for the map with the minimum number of edges mapped to each edge of the tree. This parameter is related to treewidth, but unlike treewidth, it distinguishes multigraphs from their underlying simple graphs. Stable gonality is relevant for problems in number theory.
	In this paper, we show that deciding whether the stable gonality of a given graph is at most a given integer $k$ belongs to the class NP, and we give an algorithm that computes the stable gonality of a graph in $O((1.33n)^nm^m \text{poly}(n,m))$ time. 
\end{abstract}

	\section{Introduction}

The gonality of an algebraic curve $X$ is the minimal degree of a non-constant morphism to the projective line $\PP^1$. Algorithms are known that, given equations for $X$, compute its gonality, see for example \cite{SSW}. Based on analogies between algebraic curves and graphs, various analogues of gonality have been defined in graph theory, see \cite{Baker08, Caporaso, CKK}. In this paper, we are concerned with the computation of the so-called stable gonality $\sgon(G)$ of a multigraph $G$. Stable gonality is defined as the direct analogue of the above geometric definition: $\sgon(G)$ is the minimal degree of a finite harmonic morphism of a refinement of $G$ to a tree. Here, a refinement of $G$ is given by iteratively subdividing edges or adding leaves (see \ref{def:morphism}-\ref{def:gonality} infra).
Our main result says:
\begin{introtheorem}
	There is an algorithm that, given a graph $G$, computes its stable gonality $\sgon(G)$ in time $O((1.33n)^nm^m \poly(n,m))$. Furthermore, deciding whether a graph has stable gonality at most a given integer is in NP.
\end{introtheorem}

It is not immediately clear that stable gonality of a graph is computable, since its definition involves three quantifiers over infinite sets.
In this paper we bound the number of refinements, trees and morphisms that we have to consider
and give an algorithm to compute the stable gonality of a graph,
which shows that stable gonality is computable.

There are similar notions of gonality for tropical curves and graphs, see \cite{Amini2,Caporaso}. 
Also, on algebraic curves there is an equivalent definition of gonality, using linear systems of divisors. Another notion of gonality for graphs, divisorial gonality, has been introduced using the analogue of linear systems \cite{BN07,Baker08}. Interestingly, divisorial gonality turns out to be different from stable gonality, as shown in \cite[Example 4]{Caporaso} and by an example of Luo in \cite[Example 5.13]{Amini2}. 

Some results about the computational complexity of gonality for graphs are known. For example, both notions of gonality for graphs are NP-hard to compute \cite{Gijswijt}. Deciding whether a graph has gonality 2, on the other hand, can be done in quasilinear time for both notions of gonality \cite{BBCW}. An algorithm is known that computes the divisorial gonality of a graph in $O(n^k\poly(n,m))$ time, where $n$ is the number of vertices, $m$ the number of edges and $k$ the divisorial gonality of the graph \cite{Josse}. Hence computing the divisorial gonality is in XP. 
On the other hand, it is not known whether stable or divisorial gonality can be used for fixed parameter tractable algorithms: it would be interesting to see examples of problems that are tractable on graphs of bounded gonality (either stable or divisorial gonality), but are not tractable on graphs of bounded treewidth.

Computing stable gonality is relevant in the theory of diophantine equations. More specifically, if $X$ is a smooth projective curve defined over a global field $K$ with stable reduction graph $G$ at some non-archimedean place, it is known that $\gon(X) \geq \sgon(G)$ \cite[\S 4]{CKK}. The following ``uniform boundedness result" follows: $X$ has only finitely many points in the union of all field extensions of $K$ of degree at most $(\sgon(G)-1)/2$ \cite[\S 11]{CKK}.

This paper is structured as follows. In Section \ref{sec:prelims} we introduce the definition of stable gonality. We give an algorithm to compute the stable gonality of a graph in Section \ref{sec:algorithm}, and in Section \ref{sec:boundingSize}, \ref{sec:boundingIndex} and \ref{sec:functions} we prove that this algorithm is correct. This work implies that the stable gonality problem belongs to the complexity class NP, see Section \ref{sec:NP}.

	\section{Preliminaries}
	\label{sec:prelims}

	Stable gonality is a multigraph parameter, so in this paper, we consider multigraphs; whenever we write graph, we mean finite undirected connected multigraph. A multigraph $G$ consists of a set $V(G)$ of vertices and a multiset $E(G)$ of edges. By $E_v$ we denote the set of edges incident to a vertex $v$. 
	  
	In this section, we will define stable gonality as in \cite[Definition 3.6]{CKK}, using finite harmonic morphisms. 

\begin{definition}\label{def:morphism}
	Let $G$ and $H$ be loopless graphs. A \emph{finite morphism} is a map $\phi\colon G \to H$, such that
	\begin{itemize}
		\item vertices are mapped to vertices: $\phi(v) \in V(H)$ for all $v\in V(G)$, 
		\item edges are preserved: $\phi(e) = \phi(u)\phi(v) \in E(H)$ for all $e = uv \in E(G)$.
	\end{itemize}
	together with, for every edge $e\in E(G)$, an \emph{index} $r_{\phi}(e)\in \mathbb N$.
\end{definition}

\begin{remark}
	Let $e = uv$ be an edge. Notice that $e$ is mapped to an edge connecting the images of its endpoints: $\phi(e) = e'$ with $e' =  \phi(u)\phi(v)$. Moreover, this also means that there has to be such an edge $e'$, and more specifically it holds that $\phi(u) \neq \phi(v)$. 
\end{remark}

\begin{definition}
	Let $\phi\colon G \to H$ be a finite morphism. Let $v \in V(G)$ be a vertex of $G$ and $e \in E_{\phi(v)}$ an edge in $H$.
	The \emph{index of $v$ in the direction of $e$} is
	\begin{align*}
		\sum_{d \in E_v, \phi(d) = e} r_{\phi}(d).
	\end{align*}
	We write $m_{\phi, e}(v)$ for this number. 
\end{definition}
We can think of these indices as follows: a vertex $v$ has a certain weight, namely the sum of the indices of the edges incident to $v$. The index in the direction of an edge $e$ incident to $\phi(v)$ indicates how much of this weight is send to the edge $e$. 
In order for a morphism to be harmonic, we want that every vertex distributes its weight equally over all edges incident to $\phi(v)$. 

\begin{definition}
	A finite morphism $\phi\colon G \to H$ is \emph{harmonic} if,
	for all $v\in V(G)$ and for all $e, e' \in E_{\phi(v)}$, 
	\begin{equation*}
		\sum_{d \in E_v, \phi(d) = e} r_{\phi}(d) = \sum_{d' \in E_v, \phi(d') = e' } r_{\phi}(d').
	\end{equation*}
	In other words, $\phi$ is harmonic if for every vertex the index in each direction is the same. 
	We call this number the \emph{index} of $v$ and denote it by $m_\phi(v)$. 
\end{definition}

A consequence of a finite morphism being harmonic is that the total weight that is mapped to each edge is equal. We call this amount the \emph{degree} of the finite harmonic morphism. 
\begin{definition}
	The \emph{degree} of a finite harmonic morphism $\phi\colon G \to H$ is
	\begin{equation*}
		\deg(\phi) = \sum_{d \in \phi^{-1}(e)} r_{\phi}(d)  = \sum_{u \in \phi^{-1}(v)} m_{\phi}(u), 
	\end{equation*}
	for any choice of $e\in E(H)$ or $v\in V(H)$. This number is independent of the choice \cite[Lemma 2.4]{BN07}. 
\end{definition}

\begin{definition}
	Let $G$ be a graph. A \emph{refinement} of $G$ is a graph $H$ that can be obtained by applying the following operations finitely many times:
	\begin{itemize}
		\item add a new leaf, \textit{i.e.}\ a vertex of degree $1$;
		\item subdivide an edge, \textit{i.e.}\ replace an edge by a vertex of degree $2$.
	\end{itemize}
	We call a vertex of $H\backslash G$ from which there are two disjoint paths to vertices of $G$, \emph{internal added
	vertex}, we call the other vertices of $H\backslash G$ \emph{external added vertices}.
\end{definition}

\begin{definition} \label{def:gonality}
	Let $G$ be a graph. The \emph{stable gonality} of $G$ is
	\begin{equation*}
		\begin{split}
		\sgon(G) = \min\{\deg(\phi)\mid \phi\colon H \to T \text{ a finite harmonic morphism, } \\
		H \text{ a refinement of $G$, and } T \text{ a tree}\}.
		\qedhere
		\end{split}
	\end{equation*}
\end{definition}

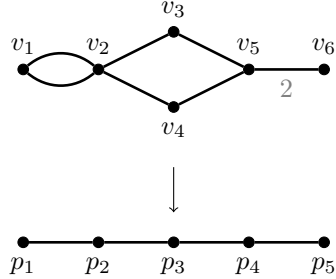
\begin{figure}
	\centering
	\begin{tikzpicture}
	\node[vertex, label=$v_1$] (a) at (0,.5) {};
	\node[vertex, label=$v_2$] (b) at (1,.5) {};
	\node[vertex, label=below:$v_4$] (c) at (2,0) {};
	\node[vertex, label=$v_3$] (d) at (2,1) {};
	\node[vertex, label=$v_5$] (e) at (3,.5) {};
	\node[vertex, label=$v_6$] (f) at (4,.5) {};
	\draw[edge] (a) to [relative,out=40, in=140] (b);
	\draw[edge] (b) to [relative,out=40, in=140] (a);
	\draw[edge] (b) -- (c) -- (e) -- (d) -- (b);
	\draw[edge] (e) -- node[below] {\color{gray} $2$} (f);
	\draw[->] (2, -.8) -- (2,-1.4);
	\node[vertex, label=below:$p_1$] (x1) at (0,-1.8) {};
	\node[vertex, label=below:$p_2$] (x2) at (1,-1.8) {};
	\node[vertex, label=below:$p_3$] (x3) at (2,-1.8) {};
	\node[vertex, label=below:$p_4$] (x4) at (3,-1.8) {};
	\node[vertex, label=below:$p_5$] (x5) at (4,-1.8) {};
	\draw[edge] (x1) -- (x2) -- (x3) -- (x4) -- (x5);
	\end{tikzpicture}
	\caption{The graph $G$ of Example \ref{ex:sgonCycle} and a finite harmonic morphism of degree $2$.} \label{fig:vbSgonCycle}
\end{figure}
\begin{figure}
	\centering
	\begin{tikzpicture}
	\node[vertex, label=$u$] (a) at (0,0) {};
	\node[vertex, label=$v$] (b) at (2,0) {};
	\draw[edge] (a) -- (b);
	\draw[edge] (a) to [relative,out=25, in=155] (b);
	\draw[edge] (b) to [relative,out=25, in=155] (a);
	\draw[edge] (a) to [relative,out=50, in=130] (b);
	\draw[edge] (b) to [relative,out=50, in=130] (a);
	\draw[->] (3,0) to (4,0);
	\node[vertex, label=$u$] (c) at (5,0) {};
	\node[vertex, label=$v$] (d) at (7,0) {};
	\draw[edge] (c) -- (d) node[midway,added] {};
	\draw[edge] (c) to [relative,out=25, in=155] (d);
	\draw[edge] (d) to [relative,out=25, in=155] (c);
	\draw[edge] (c) to [relative,out=50, in=130] (d);
	\draw[edge] (d) to [relative,out=50, in=130] (c);
	\node[added] (x1) at (6,.26) {};
	\node[added] (x2) at (6,-.26) {};
	\node[added] (x3) at (6,.48) {};
	\node[added] (x4) at (6,-.48) {};
	\draw[->] (8,0) to (9,0);
	\node[vertex, label=$v'$] (e) at (10,0) {};
	\node[added] (y) at (11,0) {};
	\node[added] (y1) at (11,.26) {};
	\node[added] (y2) at (11,-.26) {};
	\node[added] (y3) at (11,.48) {};
	\node[added] (y4) at (11,-.48) {};
	\draw[edge] (e) -- (y);
	\draw[edge] (e) to [out=25, in=180] (y1);
	\draw[edge] (e) to [out=-25, in=180] (y2);
	\draw[edge] (e) to [out=50, in=180] (y3);
	\draw[edge] (e) to [out=-50, in=180] (y4);
	\end{tikzpicture}
	\caption{The banana graph admits, after refining, a finite harmonic morphism of degree $2$. }
	\label{fig:banana}
\end{figure}
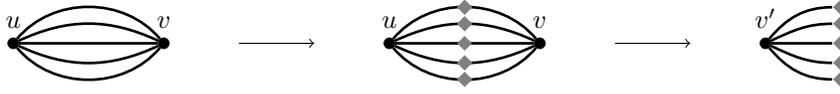

\begin{example} \label{ex:sgonCycle}
	Consider the graph in Figure \ref{fig:vbSgonCycle}. We can map this graph to the path graph on five vertices as follows: $\phi(v_1) = p_1$, $\phi(v_2) = p_2$, $\phi(v_3) = \phi(v_4) = p_3$, $\phi(v_5) = p_4$ and $\phi(v_6) = p_5$, see Figure \ref{fig:vbSgonCycle} for an illustration. Give the edge $v_5v_6$ index $2$, and all other edges index $1$. This is a finite morphism. 
	
	We can check that $\phi$ is harmonic. Consider, for example, vertex $v_5$. There are two edges incident to $\phi(v_5)$, namely $p_3p_4$ and $p_4p_5$. We can compute \begin{align*}
	m_{\phi, p_3p_4}(v_5) = \sum_{e\in E_{v_5}, \phi(e) = p_3p_4} r_\phi(e) &= r_\phi(v_3v_5) + r_\phi(v_4v_5) = 2, \\
	m_{\phi, p_4p_5}(v_5) = \sum_{e\in E_{v_5}, \phi(e) = p_4p_5} r_\phi(e) &= r_\phi(v_5v_6) = 2.
	\end{align*} We see that these sums are indeed equal, and $m_\phi(v_5) = 2$. Analogously, we can check that $m_\phi(v_1) = m_\phi(v_2) = m_\phi(v_6) = 2$ and $m_\phi(v_3) = m_\phi(v_4) = 1$. 
	
	The degree of $\phi$ is $\sum_{v\in V(G), \phi(v) = p_4} m_\phi(v) = m_\phi(v_5) = 2$. So we conclude that $\sgon(G) \leq 2$. 
\end{example}
\begin{example}[{\cite[Example 3.9]{CKK}}] \label{ex:gonBanana}
	The banana graph $B_m$ is a graph with 2 vertices $u$ and $v$ and $m\geq 2$ edges, see Figure \ref{fig:banana}. Let $\phi\colon B_m \to T$ be a finite harmonic morphism to a tree $T$. It follows that $\phi(u) \neq \phi(v)$, otherwise the edges $uv$ are not send to an edge $\phi(u)\phi(v)$. It follows that all $m$ edges are mapped to the edge $\phi(u) \phi(v)$, thus $\deg(\phi) \geq m$. 
	
	By refining $B_m$ first, we can obtain finite harmonic morphisms with lower degree. Consider the refinement $G'$ where every edge is subdivided once. Let $T$ be a tree with a vertex $v'$ and $m$ leaves, see Figure \ref{fig:banana}. Let $\phi: G' \to T$ be the map such that $\phi(u) = \phi(v) = v'$ and all other vertices are mapped to a unique leaf. Assign index $1$ to every edge of $G'$. Now we see that $\phi$ is a finite harmonic morphism of degree $2$.
\end{example}

Stable gonality is related to other graph parameters, like treewidth and the first Betti number. The first Betti number of a graph equals $m-n+1$, where $m$ is the number of edges and $n$ the number of vertices. 
Treewidth only depends on the underlying simple graph of a multigraph, but stable gonality distinguishes multigraphs and their underlying simple graphs. 
The following inequalities hold for all graphs $G$ with $m$ edges and $n$ vertices: $\tw(G) \leq \sgon(G) \leq \lfloor \frac{m-n+4}{2}\rfloor$ \cite{CKK, JosseGijswijtTreewidth}.

Lastly, we introduce notation for a part of a graph from a given vertex $v$ in the direction of a given vertex $u$. 
\begin{definition}
	Let $G$ be a graph and $u,v$ vertices.
	Let $U$ be the connected component of $G-v$ containing $u$.
	By $G_v(u)$ we denote the induced subgraph on $U\cup\{v\}$. By $G_{uv}$ we denote the graph $(G_u(v))_v(u)$. 
\end{definition}

\begin{figure}
	\centering
	\begin{tikzpicture}
	\node[vertex, label=$u$] (a) at (0,0) {};
	\node[vertex, label=$v$] (b) at (1,0) {};
	\node[vertex] (c) at (0,-1) {};
	\node[vertex] (d) at (1,-1) {};
	\node[vertex] (e) at (1.5,-1.6) {};
	\node[vertex] (f) at (-.5,-1.6) {};
	\node[vertex] (g) at (-.8,-2.3) {};
	\node[vertex] (h) at (-.2,-2.3) {};
	\node[vertex] (i) at (-1,0) {};
	\node[vertex] (j) at (1.9,0) {};
	\node[vertex] (k) at (1.8,-.5) {};
	\node[vertex] (l) at (2.5,-0.5) {};
	\draw[edge] (a) -- (b) -- (d) -- (c) -- (a);
	\draw[edge] (d) -- (e);
	\draw[edge] (c) -- (f);
	\draw[edge] (h) -- (f) -- (g);
	\draw[edge] (b) -- (j) -- (k) -- (b);
	\draw[edge] (j) -- (l) -- (k);
	\draw[edge] (a) -- (i);
	\end{tikzpicture}
	\qquad \quad
	\begin{tikzpicture}
	\node[vertex, label=$u$] (a) at (0,0) {};
	\node[vertex, label=$v$] (b) at (1,0) {};
	\node[vertex] (c) at (0,-1) {};
	\node[vertex] (d) at (1,-1) {};
	\node[vertex] (e) at (1.5,-1.6) {};
	\node[vertex] (f) at (-.5,-1.6) {};
	\node[vertex] (g) at (-.8,-2.3) {};
	\node[vertex] (h) at (-.2,-2.3) {};
	\node[vertex] (i) at (-1,0) {};
	\draw[edge] (a) -- (b) -- (d) -- (c) -- (a);
	\draw[edge] (d) -- (e);
	\draw[edge] (c) -- (f);
	\draw[edge] (h) -- (f) -- (g);
	\draw[edge] (a) -- (i);
	\end{tikzpicture}
	\qquad \quad
	\begin{tikzpicture}
	\node[vertex, label=$u$] (a) at (0,0) {};
	\node[vertex, label=$v$] (b) at (1,0) {};
	\node[vertex] (c) at (0,-1) {};
	\node[vertex] (d) at (1,-1) {};
	\node[vertex] (e) at (1.5,-1.6) {};
	\node[vertex] (f) at (-.5,-1.6) {};
	\node[vertex] (g) at (-.8,-2.3) {};
	\node[vertex] (h) at (-.2,-2.3) {};
	\draw[edge] (a) -- (b) -- (d) -- (c) -- (a);
	\draw[edge] (d) -- (e);
	\draw[edge] (c) -- (f);
	\draw[edge] (h) -- (f) -- (g);
	\end{tikzpicture}
	\caption{A graph $G$, the graph $G_v(u)$ and the graph $G_{uv}$} \label{fig:subgraphs}
\end{figure}
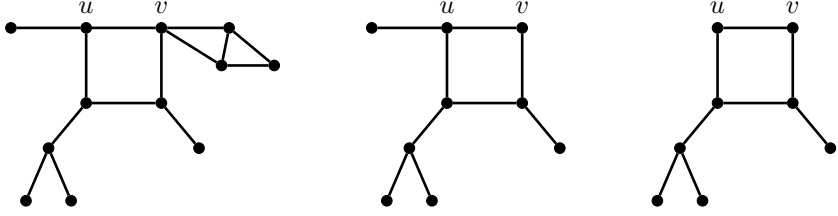

Intuitively, $G_{uv}$ is the part of $G$ `between' $u$ and $v$. See Figure \ref{fig:subgraphs} for an example. 

We say we \emph{add $G_v(u)$ to a vertex $w$ in a graph $H$}, when we add a copy of $G_v(u)$ to $H$ and identify $v$ with $w$.

We say we \emph{refine edge $uv$ of a graph $H$ as $G_{u'v'}$}, when we remove $uv$ from $H$, add a copy of $G_{u'v'}$ and identify $u$ with $u'$ and $v$ with $v'$. This can be done with any graph $G$, but in the rest of the paper, when we refine an edge as $G_{u'v'}$, the graph $G$ will always be a tree.

	\section{Algorithm overview} \label{sec:algorithm}
	In this paper we will give an algorithm to compute the stable gonality of a graph. Notice that this is not initially trivial: there are infinitely many refinements of a graph, there are infinitely many trees and there are infinitely many finite harmonic morphisms, since there are infinitely many assignments of indices to the edges. We bound the trees and maps that we have to consider; our algorithm enumerates all those trees and maps. 

	The algorithm considers all tuples $\alpha = (T,f,r)$, where
	\begin{itemize}
		\item $T$ is a tree with at most $n=|V(G)|$ vertices,
		\item $f:V(G)\to V(T)$ is a surjective map,
		\item $r:E(G)\to [\lfloor\frac{m-n+4}{2}\rfloor]$, where we denote $[k]$ for the set $\{1, 2, 3, \ldots, k\}$,
		is a map assigning indices to the edges of $G$.
	\end{itemize}
	Given such a tuple, we construct a finite harmonic morphism $\phi_{\alpha}$ from a refinement of $G$ to
	a tree constructed from $T$ by optionally adding at most $m$ leaves. We compute the degree of $\phi_\alpha$, and output the minimum degree over all tuples $\alpha$. 
	The remainder of this section covers the construction of $\phi_\alpha$
	and analyses the runtime of the algorithm.
	The remainder of the paper proves that the assumptions made by the algorithm are valid.
	Sections \ref{sec:boundingSize} and \ref{sec:boundingIndex} show that it suffices to only consider
	trees $T$ of size at most $n$ and indices $r_\phi(e)$ at most $\floor{(m-n+4)/2}$. 
	Section \ref{sec:functions} proves that there exists a finite harmonic morphism of the form $\phi_\alpha$ that
	attains the minimal degree $\sgon(G)$ indeed.

	\subsection{Construction of $\phi_\alpha$}\label{subsec:construction}
	We now explain how to construct a refinement $H$
	and a finite harmonic morphism $\phi_\alpha$ from a tuple $\alpha = (T,f,r)$.

\begin{figure}
	\centering
	\begin{tikzpicture}
	\node[vertex, label=below:$v$] (u) at (0,0) {};
	\node[vertex, label=$w$] (v) at (0,1) {};
	\node[vertex, label=$x$] (w) at (0,3) {};
	\node[vertex, label=left:$u$] (x1) at (-1,1) {};
	\node[vertex, label=right:$y$] (x2) at (1,1) {};
	\draw[edge] (u) -- (v); 
	\draw[edge] (x1) to[in=130, out=50] (x2);

	\draw[->] (0,-.7) -- (0,-1.3);

	\node[vertex] (v') at (0,-2) {};
	\node[vertex] (x) at (-1,-2) {};
	\node[vertex] (x') at (1,-2) {};
	\draw[edge] (x) -- (v')--(x');
	\draw[edge] (x1) -- (u) -- (x2);
	\draw[edge] (x1) -- (v) -- (x2) -- (w) -- (x1);

	\draw[->] (1.7,0) -- (2.3,0);

	\node[vertex, label=below:$v$] (u) at (4,0) {};
	\node[vertex, label=$w$] (v) at (4,1) {};
	\node[vertex] (w) at (4,3) {};
	\node[vertex] (x1) at (3,1) {};
	\node[vertex] (x2) at (5,1) {};
	\draw[edge] (u) -- (v); 
	\draw[edge] (x1) to[in=130, out=50] (x2);

	\draw[->] (4,-.7) -- (4,-1.3);

	\node[vertex] (v') at (4,-2) {};
	\node[vertex] (x) at (3,-2) {};
	\node[vertex] (x') at (5,-2) {};
	\draw[edge] (x) -- (v')--(x');
	\draw[edge] (x1) -- (u) -- (x2);
	\draw[edge] (x1) -- (v) -- (x2) -- (w) -- (x1);
	\node[added, label={[label distance=-5pt]30:{\small$e_{vw}$}}] (e) at (4,.5) {};
	\node[added, label=right:$e'_{vw}$] (e') at (4,-2.5) {};
	\draw[edge] (v') -- (e');

	\draw[->] (5.7,0) -- (6.3,0);

	\node[vertex] (u) at (8,0) {};
	\node[vertex] (v) at (8,1) {};
	\node[vertex] (w) at (8,3) {};
	\node[vertex, label=left:$u$] (x1) at (7,1) {};
	\node[vertex, label=right:$y$] (x2) at (9,1) {};
	\draw[edge] (u) -- (v); 
	\node[added] (y) at (8,1.5) {};
	\node[added] (y') at (8,2) {};
	\draw[edge] (y) -- (y');
	\node[added] (e) at (8,.5) {};
	\draw[edge] (x1) -- (u) -- (x2);
	\draw[edge] (x1) -- (v) -- (x2) -- (w) -- (x1);
	\draw[edge] (x1) to[in=183,out=50] (y);
	\draw[edge] (x2) to[in=-03,out=130] (y);

	\draw[->] (8,-.7) -- (8,-1.3);

	\node[vertex] (v') at (8,-2) {};
	\node[vertex] (x) at (7,-2) {};
	\node[vertex] (x') at (9,-2) {};
	\draw[edge] (x) -- (v')--(x');
	\node[added] (e') at (8,-2.5) {};
	\draw[edge] (v') -- (e');

	\draw[->] (9.7,0) -- (10.3,0);

	\node[vertex] (u) at (12,0) {};
	\node[vertex] (v) at (12,1) {};
	\node[vertex, label=right:$x$] (w) at (12,3) {};
	\node[vertex] (x1) at (11,1) {};
	\node[vertex] (x2) at (13,1) {};
	\draw[edge] (u) -- (v); 
	\node[added] (y) at (12,1.5) {};
	\node[added] (y') at (12,2) {};
	\draw[edge] (y) -- (y');
	\node[added,label=right:$l_x$] (lw) at (12,3.5) {};
	\node[added] (e) at (12,.5) {};
	\draw[edge] (w) -- (lw);
	\draw[edge] (x1) -- (u) -- (x2);
	\draw[edge] (x1) -- (v) -- (x2) -- (w) -- (x1);
	\draw[edge] (x1) to[in=183,out=50] (y);
	\draw[edge] (x2) to[in=-03,out=130] (y);

	\draw[->] (12,-.7) -- (12,-1.3);

	\node[vertex] (v') at (12,-2) {};
	\node[vertex] (x) at (11,-2) {};
	\node[vertex] (x') at (13,-2) {};
	\draw[edge] (x) -- (v')--(x');
	\node[added] (e') at (12,-2.5) {};
	\draw[edge] (v') -- (e');
	\end{tikzpicture}
	\caption{
		Consider the map where every vertex is mapped to the vertex of the tree below it. 
		When both ends of an edge are mapped to the same vertex, for example edge $vw$,
		we add a vertex $e_{vw}$ and map it to a new vertex $e'_{vw}$.
		Then, we refine edges for which the ends are not mapped to the same vertex, like edge $uy$, as the part of the tree they correspond with.
		Lastly, we add copies of part of the tree, like $l_x$, to make sure the morphism is harmonic at every vertex.
		In this example, all edges have index $1$.
	}
	\label{fig:verfijning1}
\end{figure}
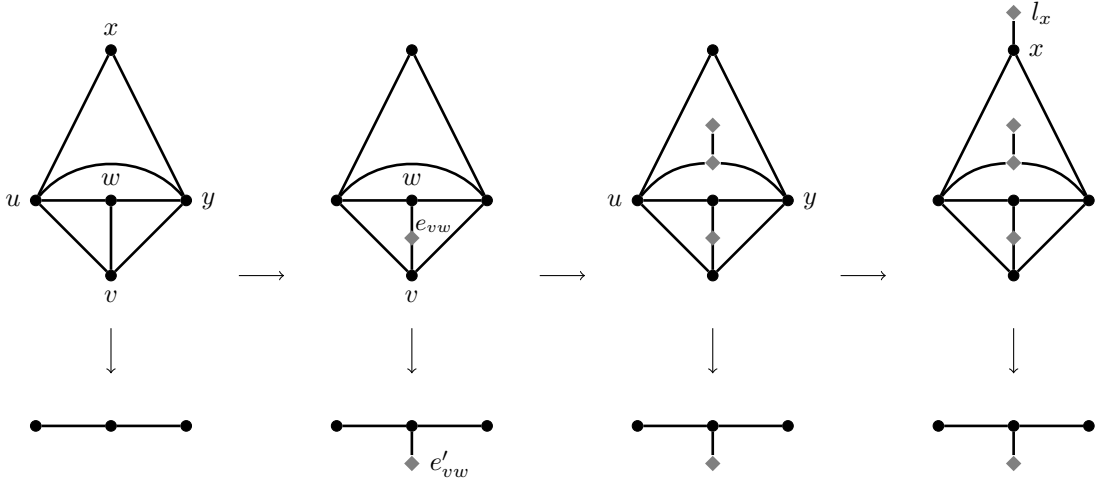

	First we set $\phi(v) = f(v)$ for every vertex $v$ of $G$.
	For each edge $uv \in E(G)$ with $f(u) = f(v)$, we add a vertex $e_{uv}$ to $uv$.
	Besides, we add a leaf $e'_{uv}$ to $f(u)$.
	We assign index 1 to those new edges and set $\phi(e_{uv}) = e'_{uv}$.
	This is depicted in the second column of Figure \ref{fig:verfijning1}.
	Write $T'$ for the tree we constructed from $T$ by adding these leaves. 
	Now, for every edge $uv\in E(G)$ with $f(u)\neq f(v)$, we refine $uv$ as $T'_{f(u)f(v)}$.
	Assign index $r(uv)$ to all those new edges and use the identity map to map this part of the refinement to $T'_{f(u)f(v)}$.
	This is the third column in the figure.
	In the last column, we ensure that our map is harmonic: for every vertex $v\in V(G)$,
	let $e \in E_{f(v)}$ be the edge such that $m_{\phi, e}(v)$ is maximal.
	Now, for every edge $e' = f(v)u' \in E_{f(v)}$ with $m_{\phi, e'}(v) < m_{\phi, e}(v)$,
	we add $T'_{f(v)}(u')$ to $v$,
	assign index $m_{\phi, e}(v) - m_{\phi, e'}(v)$ to these new edges and use the identity map to map this
	part of the refinement to $T'_{f(v)}(u')$.
	
	Let $H_\alpha$ be the refinement constructed in this way, and set $\phi_\alpha = \phi$. Now $\phi_\alpha$ is a finite harmonic morphism from $H_\alpha$ to $T'$. 

\subsection{Runtime analysis}

We now analyse the runtime of the algorithm.
We first count the number of pairs $(T,f)$ where $T$ is a tree and $f$ is a surjective map from $G$ to $T$.

By Cayley's formula, there are $k^{k-2}$ labelled trees of size $k$ \cite[Section 2.3.4.4]{KnuthTAOCP}.
The number of unlabelled partitions of the $n$ vertices of $G$ into $k$ non-empty sets is
$\stirling nk$, the Stirling number of the second kind \cite[Section 1.2.6]{KnuthTAOCP}.
By assigning one label to each set, we see that the number of surjective maps to a fixed tree of size $1\leq
k\leq n$ is $\stirling nk$ indeed.
Summing over all trees of size at most $n$, the total number of pairs $(T,f)$ is
\begin{equation*}
	\sum_{k=1}^n k^{k-2} \stirling nk.
\end{equation*}
We will bound this quantity.
When we first choose one vertex for each set,
and then distribute the remaining $n-k$ vertices amongst those sets,
we overcount the number of unlabelled partitions.
Hence, $\stirling nk \leq \binom nk k^{n-k}$.
This implies
\begin{equation*}
\sum_{k=1}^n k^{k-2} \stirling nk
\leq \sum_{k=1}^{n} k^{k-2} \binom nk k^{n-k}
\leq \sum_{k=1}^{n} k^n \binom nk.
\end{equation*}
By Stirling's approximation we have $\binom{n}{k} \leq (\frac{n}{k})^k(\frac{n}{n-k})^{n-k}$ \cite[Section 1.2.6]{KnuthTAOCP}. 
We infer that
\begin{equation*}
\sum_{k=1}^n k^{k-2} \stirling nk \leq
\sum_{k=1}^{n} k^{n} \cdot n^n \cdot  k^{-k} \cdot (n-k)^{-(n-k)}
= \sum_{k=1}^{n} n^n \cdot  \parens*{\frac k{n-k}}^{n-k}.
\end{equation*}
Now write $x$ for $k/n$. We have
\begin{equation*}
\parens*{\frac k{n-k}}^{(n-k)/n} = \parens*{\frac{x}{1-x}}^{1-x}.
\end{equation*}
Since this has a finite limit in both $x\to 0$ and $x\to 1$,
we may consider the maximum on the $x\in (0,1)$ interval.
A calculation shows that this maximum is less than $1.33$. 
Substituting this in the bound on $\sum_{k=1}^n k^{k-2} \stirling nk$ yields
\begin{equation*}
\sum_{k=1}^n k^{k-2} \stirling nk
\leq \sum_{k=1}^n n^n\cdot 1.33^n
\leq n\cdot (1.33n)^n
\leq (1.33n)^{n+1}.
\end{equation*}

The number of functions $r:E(G)\to \brackets*{\floor*{\frac{m-n+4}2}}$ assigning indices to the edges is
$ \floor*{\frac{m-n+4}2}^m $.

When we are given a tuple $\alpha = (T,f,r)$, the construction of $H$ and $\phi_\alpha$ can be done in
polynomial time in $n$ and $m$: we have to consider every edge once to refine it and we have to consider every vertex once to make the map harmonic. The calculation of the degree can also be done in polynomial time, by picking an edge $e$ in the tree, and check for every edge of $H$ whether it is mapped to $e$.
It follows that the runtime of our algorithm is bounded by
\begin{equation*}
	O\parens*{\poly(n,m) \cdot (1.33n)^n \parens*{\frac{m-n+4}2}^m}.
\end{equation*}

\begin{remark}
	In practice, it can be useful to do some pre-processing first, although this does not change the runtime in general.
	Before trying all tuples $\alpha$, we can
	contract all vertices of degree one or two.
	The graph obtained in this way is called a \emph{stable graph}. 
	It is known that this stable graph has the same stable gonality
	as the graph we started with \cite[Lemma 5.4]{CKK}.
\end{remark}	
\begin{remark}
	A C++ implementation of this algorithm is made. Unfortunately, this algorithm is too time consuming to use for graphs with more than 5 vertices. The implementation is available from the first author upon request. 
\end{remark}

	\section{Bounding the size of the tree} \label{sec:boundingSize}
	
	In this section we will show that we only have to consider finite harmonic morphisms to trees with at most
	$|V(G)|$ internal (\textit{i.e.}\ non leaf) vertices.
	In particular, we show that any finite harmonic morphism can be transformed in such a way
	that every internal vertex of $T$ is covered by at least one vertex of $G$.

	\begin{definition}
		Let a graph $G$, a refinement $H$ of $G$, a tree $T$,
		and a finite harmonic morphism $\phi \colon H\to T$ be given.
		A \emph{transformation} of $\phi$ is a new finite harmonic morphism $\phi' \colon H'\to T'$ where
		$H'$ is again a refinement of $G$ and $T'$ a tree, such that $\deg(\phi') \leq \deg(\phi)$.
	\end{definition}
	\begin{remark}
		From here on $H$ and $H'$ will always be refinements of $G$ while $T$ and $T'$ will always be trees.
		The phrase \emph{let $\phi\colon H\to T$ be given} will implicitly assume this.
	\end{remark}
	
	\begin{definition}
		Let $G$ be a graph with a refinement $H$.
		A vertex $v\in V(H)$ is \emph{alien} if it is not a vertex in $G$.
		Let $\phi\colon H\to T$ be a finite harmonic morphism.
		A vertex $v'\in V(T)$ is \emph{alien} if all vertices in $\phi^{-1}(v')$ are alien. 
	\end{definition}

	The main lemma of this section is that we can remove all internal alien vertices from $T$.
	\begin{lemma} \label{lem:noInternalAliens}
		Let $\phi\colon H\to T$ be given.
		There exists a transformation $\phi' \colon H' \to T'$ such that $T'$ has no internal alien vertices.
	\end{lemma}

	To prove this 
	we contract all internal alien vertices in $T$ to a neighbour,
	while simultaneously contracting all edges in the preimages of these edges. By $G/e$ we denote the graph obtained by contracting edge $e$ in graph $G$.

	\begin{lemma} \label{lem:contractingEdges}
		Let $\phi \colon H \to T$ be given, and let $v'$ be an internal alien vertex of $T$ with a neighbour $u'$.
		There exists a transformation $\phi' \colon H' \to T/u'v'$.
	\end{lemma}

	\begin{proof}
		Let $v_1, \ldots, v_k$ be the vertices of $H$ that are mapped to $v' $ and let $u_1, \ldots, u_l$ be
		the vertices that are mapped to $u'$. 
		We will construct a refinement $H'$ of $G$ as follows. 
		Consider a vertex $v:=v_i$.
		Notice that $v$ is an alien vertex of $H$, so there are at most two edges $vw_{1}$
		and $vw_{2}$ such that $H_{v}(w_{1})$ and $H_{v}(w_{2})$ contain vertices of $G$.

		\case{Case 1}
		If there is at most one vertex $u_j$ neighbouring $v$ such that $H_{v}(u_j)$ contains vertices of $G$,
		then contract all edges $vu_h$, and keep all indices.
		This process is shown in Figure \ref{fig:treesize}.

		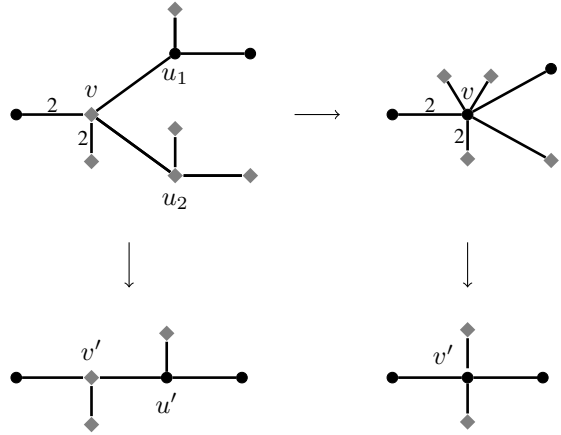
\begin{figure}
	\centering
	\begin{tikzpicture}
		\node[added, label=$v$] (v) at (0,3) {};
		\node[vertex,left of=v] (x) {};
		\node[vertex,above right=0.7cm and 1cm of v,label=below:$u_1$] (u1) {};
		\node[added,below right=0.7cm and 1cm of v,label=below  :$u_2$] (u2) {};
		\node[vertex,right of=u1] (y1) {};
		\node[added,right of=u2] (y2) {};
		\node[added,below=0.4cm of v] (l1) {};
		\node[added,above=0.4cm of u1] (l21) {};
		\node[added,above=0.4cm of u2] (l22) {};
		\draw[edge] (x) -- node[label={[label distance=-0.23cm]90:\footnotesize 2}]{} (v) -- node[label={[label distance=-0.23cm]180:\footnotesize 2}]{} (l1);
		\draw[edge] (y1) -- (u1) -- (v) -- (u2) -- (y2);
		\draw[edge] (u1) -- (l21) (v) -- (u2) -- (y2);
		\draw[edge] (u1) -- (l21);
		\draw[edge] (u2) -- (l22);

		\draw[->] (0.5,1.3) -- (0.5,0.7);

		\node[added, label=$v'$] (v') at (0,-0.5) {};
		\node[vertex,left of=v'] (x') {};
		\node[vertex,right of=v',label=below:$u'$] (u') {};
		\node[vertex,right of=u'] (y') {};
		\draw[edge] (x') -- (v') -- (u') -- (y');
		\node[added,below=0.4cm of v'] (l1') {};
		\node[added,above=0.4cm of u'] (l2') {};
		\draw[edge] (v') -- (l1');
		\draw[edge] (u') -- (l2');

		\draw[->] (2.7,3) -- (3.3,3);

		\node[vertex,label=$v$] (v) at (5,3) {};
		\node[vertex,left of=v] (x) {};
		\node[vertex,above right=0.5cm and 1cm of v] (y1) {};
		\node[added, below right=0.5cm and 1cm of v] (y2) {};
		\node[added,below=0.4cm of v] (l1) {};
		\node[added,above left =0.4cm and 0.2cm of v] (l21) {};
		\node[added,above right=0.4cm and 0.2cm of v] (l22) {};
		\draw[edge] (x) -- node[label={[label distance=-0.23cm]90:\footnotesize 2}]{} (v) -- node[label={[label distance=-0.23cm]180:\footnotesize 2}]{} (l1);
		\draw[edge] (y1) -- (v) -- (y2);
		\draw[edge] (l21) -- (v) -- (l22);

		\draw[->] (5,1.3) -- (5,0.7);

		\node[vertex,label=above left:$v'$] (v') at (5,-0.5) {};
		\node[vertex,left of=v'] (x') {};
		\node (u') at (v') {};
		\node[vertex,right of=u'] (y') {};
		\draw[edge] (x') -- (v') -- (u') -- (y');
		\node[added,below=0.4cm of v'] (l1') {};
		\node[added,above=0.4cm of u'] (l2') {};
		\draw[edge] (v') -- (l1');
		\draw[edge] (u') -- (l2');

	\end{tikzpicture}
	\caption{
		In the first case of the proof of Lemma \ref{lem:contractingEdges},
		we contract vertex $v$ to all its neighbours $u_j$ mapping to $u'$.
	}
	\label{fig:treesize}
\end{figure}

		\case{Case 2}
		Now suppose that there are two vertices $u_{j_1}$ and $u_{j_2}$ that are neighbours of $v$ such that
		$H_{v}(u_{j_1})$ and $H_{v}(u_{j_2})$ contain vertices of $G$.
		Since $v'$ is internal, there is a neighbour $x'$ of $v'$ not equal to $u'$.
		Contract $v$ to all its neighbours that map to $x'$.
		Then, for each neighbour $y$ of $v$ that maps to a vertex different from $x'$ and $u'$,
		remove the tree $G_{v}(y)$ from $H$ and
		add a copy of $T_{v'}(\phi(y))$ to each of the $u_j$.
		We assign index $r_\phi(vu_j)$ to the trees that we just added to $u_j$,
		while all other edges keep their current index.
		This is demonstrated in Figure \ref{fig:treesize2}.

		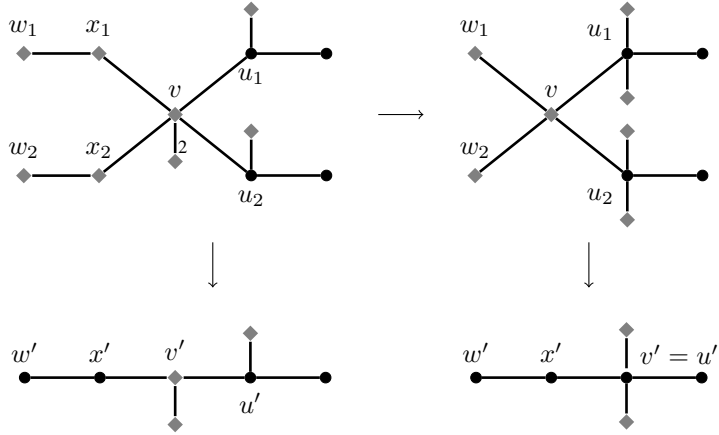
\begin{figure}
	\centering
	\begin{tikzpicture}
		\node[added, label=$v$] (v) at (0,3) {};
		\node[added,above left=0.7cm and 0.9cm of v,label=$x_1$] (x1) {};
		\node[added,below left=0.7cm and 0.9cm of v,label=$x_2$] (x2) {};
		\node[added,left of=x1,label=$w_1$] (w1) {};
		\node[added,left of=x2,label=$w_2$] (w2) {};
		\node[vertex,above right=0.7cm and 0.9cm of v,label=below:$u_1$] (u1) {};
		\node[vertex,below right=0.7cm and 0.9cm of v,label=below:$u_2$] (u2) {};
		\node[vertex,right of=u1] (y1) {};
		\node[vertex,right of=u2] (y2) {};
		\node[added,below=0.4cm of v] (l1) {};
		\node[added,above=0.4cm of u1] (l21) {};
		\node[added,above=0.4cm of u2] (l22) {};
		\draw[edge] (w1) -- (x1) -- (v) -- (x2) -- (w2);
		\draw[edge] (v) -- node[label={[label distance=-0.23cm]0:\scriptsize 2},style={pos=0.8}] {} (l1);
		\draw[edge] (y1) -- (u1) -- (v) -- (u2) -- (y2);
		\draw[edge] (u1) -- (l21) (u2) -- (l22);

		\draw[->] (0.5,1.3) -- (0.5,0.7);

		\node[added, label=$v'$] (v') at (0,-0.5) {};
		\node[vertex,left of=v',label=$x'$] (x') {};
		\node[vertex,left of=x',label=$w'$] (w') {};
		\node[vertex,right of=v',label=below:$u'$] (u') {};
		\node[vertex,right of=u'] (y') {};
		\node[added,below=0.4cm of v'] (l1') {};
		\node[added,above=0.4cm of u'] (l2') {};
		\draw[edge] (w') -- (x') -- (v') -- (u') -- (y');
		\draw[edge] (v') -- (l1');
		\draw[edge] (u') -- (l2');

		\draw[->] (2.7,3) -- (3.3,3);

		\node[added, label=$v$] (v) at (5,3) {};
		\node[added,above left=0.7cm and 0.9cm of v,label=$w_1$] (w1) {};
		\node[added,below left=0.7cm and 0.9cm of v,label=$w_2$] (w2) {};
		\node[vertex,above right=0.7cm and 0.9cm of v,label=above left:$u_1$] (u1) {};
		\node[vertex,below right=0.7cm and 0.9cm of v,label=below left:$u_2$] (u2) {};
		\node[vertex,right of=u1] (y1) {};
		\node[vertex,right of=u2] (y2) {};
		\node[added,below=0.4cm of u1] (l11) {};
		\node[added,below=0.4cm of u2] (l12) {};
		\node[added,above=0.4cm of u1] (l21) {};
		\node[added,above=0.4cm of u2] (l22) {};
		\draw[edge] (w1) -- (v) -- (w2);
		\draw[edge] (y1) -- (u1) -- (v) -- (u2) -- (y2);
		\draw[edge] (l11) -- (u1) -- (l21) (l12) -- (u2) -- (l22);

		\draw[->] (5.5,1.3) -- (5.5,0.7);

		\node[vertex,label=above right:${v'=u'}$] (v') at (6,-0.5) {};
		\node[vertex,left of=v',label=$x'$] (x') {};
		\node[vertex,left of=x',label=$w'$] (w') {};
		\node (u') at (v') {};
		\node[vertex,right of=u'] (y') {};
		\node[added,below=0.4cm of v'] (l1') {};
		\node[added,above=0.4cm of u'] (l2') {};
		\draw[edge] (w') -- (x') -- (v') -- (u') -- (y');
		\draw[edge] (v') -- (l1');
		\draw[edge] (u') -- (l2');

	\end{tikzpicture}
	\caption{
		In the second case of the proof of Lemma \ref{lem:contractingEdges},
		we contract vertex $v$ to all its neighbours $x_i$ mapping to $x'\neq u'$.
		We also remove all other trees from $v$, and copy them to all $u_j$.
	}
	\label{fig:treesize2}
\end{figure}

		Repeat this for all vertices $v_i$.
		Write $H'$ for the resulting graph.

		In the construction of $H'$, we never contract two vertices of $G$ to the same vertex,
		so $H'$ is a refinement of $G$.
		Now consider the map $\phi' \colon H' \to  T/u'v'$ we constructed.
		This map is a finite harmonic morphism with degree at most the degree of $\phi$.
	\end{proof}

	\begin{proof}[Proof of Lemma \ref{lem:noInternalAliens}]
		Repeatedly apply Lemma \ref{lem:contractingEdges}. 
	\end{proof}

	It follows that, for any given morphism $\phi\colon H \to T$, there is a transformation $\phi'\colon H' \to T'$ such that $\phi'(V(G))$ is a connected subtree of $T$ with at most $n$ vertices. So, for our algorithm it suffices to only consider trees of size at most $n$ and surjective maps to those trees.

	\section{Bounding the indices} \label{sec:boundingIndex}
	
	Cornelissen et al.\ \cite[Theorem 5.7]{CKK} gave an upper bound on the stable gonality of a graph $G$:
	\begin{equation*}
		\sgon(G) \leq \left\lfloor \frac{m-n+4}{2} \right\rfloor.
	\end{equation*}
	From this it follows that all finite harmonic morphisms $\phi\colon H\to T$,
	from a refinement $H$ of $G$ to a tree $T$ with $\deg(\phi) = \sgon(G)$,
	assign index at most $\lfloor \frac{m-n+4}{2} \rfloor$ to the edges of $H$.
	Hence it is sufficient for our algorithm to only consider functions $r\colon E(G) \to [\lfloor \frac{m-n+4}{2} \rfloor]$.

	\section{Reduction to $\phi_\alpha$}\label{sec:functions}
	
	In this section, we will prove that our algorithm will find a finite harmonic morphism of minimal degree. Let $G$ be a graph. We know that there exists a refinement $H$ of $G$, a tree $T$ and a finite harmonic morphism $\phi$ of degree $\sgon(G)$. We will show that we can transform $\phi$ to a morphism $\phi_\alpha$ for some tuple $\alpha$. In all lemmas we assume $G$ to be the given graph, $H$ a refinement, $T$ a tree and $\phi\colon H\to T$ a finite harmonic morphism of degree $\sgon(G)$.

	First, we will prove that, when two vertices $u,v$ are mapped to the same vertex, we can refine the edges $uv$ by adding just one vertex. 
	For ease of notation we give the following two graphs names: we write $P_3$ for the path on three vertices
	and $C_2$ for the cycle of length two. 
	
	\begin{lemma}\label{lem:refinementLoop}
		Let $\phi\colon H \to T$ be given.
		Let $uv \in E(G)$ be an edge such that $\phi(u) = \phi(v)$.
		There exists a transformation $\phi' \colon H' \to T'$ such that $R'_{uv} \in \{P_3, C_2\}$,
		where $R'_{uv}$ is the refinement of $uv$ in $H'$. 
	\end{lemma}
	\begin{proof}
		Let $R_{uv}$ be the refinement of the edge $uv$ in $H$.
		Let $x$ be the neighbour of $u$ in $R_{uv}$ and let $y$ be the neighbour of $v$.
		Replace $R_{uv}$ by $P_3$ when $u\neq v$, and by $C_2$ when $u=v$.
		Write $w$ for the new vertex.
		Assign index $1$ to the two new edges. 
		If $m_\phi(u) > 1$, add a leaf $w_1$ to $u$ and assign index $m_\phi(u) -1$ to it. 
		Do the same for $v$.
		For all vertices $z\in \phi^{-1}(\phi(v))\backslash R_{uv}$,
		add a leaf $w_z$ to $z$ and assign index $m_\phi(z)$ to the edge $zw_z$.
		Add a copy of $T_{\phi(u)}(\phi(x))$ to $u$ and assign index $r_\phi(ux)$ to all edges in this tree.
		Add a copy of $T_{\phi(v)}(\phi(y))$ to $v$ and assign index $r_\phi(vy)$ to the edges.
		Write $H'$ for this refinement of $G$.
		Add a vertex $w'$ to $\phi(v)$ in $T$ and write $T'$ for this new tree.
		Let $\phi' \colon H' \to T'$ be the map that sends all vertices of $z\in V(H)$ to $\phi(z)$, that sends $w, w_1, w_2$ and all vertices $w_z$ to $w'$ and uses the identity map to send the vertices of the trees we added.
		Notice that $\phi'$ is a finite harmonic morphism with $\deg(\phi')\leq \deg(\phi)$. 
	\end{proof}

	Second, we will prove that for two vertices $u,v$ that are not mapped to the same vertex, we can refine the edges $uv$ as $T_{\phi(u)\phi(v)}$. 
	\begin{lemma}\label{lem:refinementEdge}
		Let $\phi\colon H \to T$ be given. Let $uv \in E(G)$ be an edge such that $\phi(u) \neq \phi(v)$. There exists a transformation $\phi' \colon H' \to T$ such that $R'_{uv} = T_{\phi(u)\phi(v)}$, where $R'_{uv}$ is the refinement of $uv$ in $H'$. 
	\end{lemma}
	\begin{proof}
		Let $R_{uv}$ be the refinement of the edge $uv$ in $H$. Let $x$ be the neighbour of $u$ in $R_{uv}$ and let $y$ be the neighbour of $v$. We distinguish three cases. 
		
		First suppose that $\phi(x) \in T_{\phi(u)\phi(v)}$ and $\phi(y) \in T_{\phi(u)\phi(v)}$. 
		Replace $R_{uv}$ by $T_{\phi(u)\phi(v)}$ in $H$. 
		Assign index $\min\{r_\phi(ux), r_\phi(vy)\}$ to the new edges. 
		Assume, without loss of generality, that $r_\phi(ux) \leq r_\phi(vy)$. 
		If $r_\phi(ux) < r_\phi(vy)$, add a copy of the tree $T_{\phi(u)}(\phi(x))$ to $u$ and assign index $r_\phi(vy) - r_\phi(ux)$ to the new edges. 
		Write $H'$ for this new refinement of $G$. 
		Let $\phi' \colon H' \to T$ be the map that sends all vertices of $w\in V(H)$ to $\phi(w)$ and uses the identity map to send the new vertices to $T_{\phi(u)\phi(v)}$ and $T_{\phi(u)}(\phi(x))$. 
		Notice that $\phi'$ is a finite harmonic morphism with $\deg(\phi') \leq \deg(\phi)$. 
		
		Now suppose that $\phi(x) \notin T_{\phi(u)\phi(v)}$ and $\phi(y) \notin T_{\phi(u)\phi(v)}$. Replace $R_{uv}$ by $T_{\phi(u)\phi(v)}$ in $H$. Assign index $1$ to the new edges. Add a copy of $T_{\phi(u)}(\phi(x))$ to $u$ and assign index $r_\phi(ux)+ 1$ to its edges. For all neighbours $z$ of $\phi(u)$ such that $z\neq \phi(x)$ and $z \notin T_{\phi(u)\phi(v)}$, add a copy of $T_{\phi(u)}(z)$ to $u$ and assign index 1 to the new edges. Do the same for vertex $v$. Write $H'$ for this refinement of $G$. Let $\phi'\colon H' \to T$ be the map that sends all vertices $w\in V(H)$ to $\phi(w)$ and uses the identity map for all new vertices. Notice that by the choice of the indices $\phi'$ is a finite harmonic morphism with $\deg(\phi') \leq \deg(\phi)$. 
		
		In the last case, suppose, without loss of generality,
		that $\phi(x) \in T_{\phi(u)\phi(v)}$ and $\phi(y)\notin T_{\phi(u)\phi(v)}$.
		Replace $R_{uv}$ by $T_{\phi(u)\phi(v)}$ in $H$.
		Assign index $1$ to the new edges.
		Add a copy of $T_{\phi(v)}(\phi(y))$ to $v$ and assign index $r_\phi(vy)+ 1$ to its edges.
		For all neighbours $z$ of $\phi(v)$ such that $z\neq \phi(y)$ and $z \notin T_{\phi(u)\phi(v)}$, add a copy of $T_{\phi(v)}(z)$ to $v$ and assign index 1 to the new edges. 
		If $r_\phi(ux) > 1$, add a copy of $T_{\phi(u)}(\phi(x))$ to $u$ and assign index $r_\phi(ux) - 1$ to its edges.
		Write $H'$ for this refinement of $G$.
		Let $\phi'\colon H' \to T$ be the map that sends all vertices $w\in V(H)$ to $\phi(w)$ and uses the identity map for all new vertices.
		Notice that by the choice of the indices $\phi'$ is a finite harmonic morphism with $\deg(\phi') \leq \deg(\phi)$. 
	\end{proof}

	Lastly, we prove two lemmas that ensure that there are not more external added vertices than necessary. 
	\begin{lemma} \label{lem:refinementNoExternalLeaf}
		Let $\phi\colon H \to T$ be given.
		Let $v' \in V(T)$ be a leaf such that
		all vertices $v\in V(H)$ with $\phi(v) = v'$ are external added vertices.
		Define $T' = T\backslash\{v'\}$.
		There exists a transformation $\phi' \colon H' \to T'$. 
	\end{lemma}
	\begin{proof}
		Remove all vertices that are mapped to $v'$ from $H$. If $H$ becomes disconnected, remove all connected components that do not contain a vertex of $G$. Since all removed vertices were external added vertices, there is only one remaining component. Write $H'$ for this graph. Define $\phi'\colon H' \to T'$ as the restriction of $\phi$ to $H'$. Notice that $\phi'$ is a finite harmonic morphism, and $\deg(\phi') \leq \deg(\phi)$. 
	\end{proof}

	\begin{lemma} \label{lem:refinementAtMostOneAddedTree}
		Let $\phi\colon H \to T$ be given. Let $v$ be a vertex of $G$ and $u'$ a neighbour of $\phi(v)$. There exists a transformation $\phi' \colon H' \to T$ such that $v$ has at most one external added neighbour $u$ such that $\phi'(u) = u'$, and moreover, $G_v(u) = T_{\phi(v)}(u')$. 
	\end{lemma}
	\begin{proof}
		Let $u_1, \dots, u_l \in $ be the external added neighbours of $v$ such that $\phi(u_i) = u'$. Remove all trees $G_v(u_i)$ from $G$ and add one copy of $T_{\phi(v)}(u')$ to $v$. Assign index $\sum_{i=1}^{l} r_\phi(vu_i)$ to all new edges. Notice that $\phi'$ is a finite harmonic morphism, and $\deg(\phi') \leq \deg(\phi)$. 
	\end{proof}

	\begin{lemma}\label{lem:morphismProperties}
		Let $G$ be a graph with $\sgon(G) =k$. Then there is a refinement $H$ of $G$, a tree $T$ and a finite harmonic morphism $\phi\colon H\to T$ of degree $k$, that satisfy the following properties:
		\begin{itemize}
			\item $T$ does not contain internal alien vertices.
			\item For  every edge $uv\in E(G)$ such that $\phi(u) = \phi(v)$, it holds that $R_{uv} \in \{P_3, C_2\}$, where $R_{uv}$ is the refinement of $uv$ in $H$; moreover, the two edges of $R_{uv}$ have index $1$.
			\item For every edge $uv\in E(G)$ such that $\phi(u) \neq \phi(v)$, it holds that $R_{uv} = T_{\phi(u)\phi(v)}$, where $R_{uv}$ is the refinement of $uv$ in $H$; moreover, every edge in $R_{uv}$ is assigned the same index.
			\item For every vertex $v' \in V(T)$, let $v_1, \ldots, v_l$ be the vertices of $H$ such that $\phi(v_i) = v'$, then, not all of $v_1, \ldots, v_l$ are external added vertices. 
			\item For every vertex $v\in V(G)$ and for every neighbour $u'$ of $\phi(v)$, $v$ has at most one external added neighbour $u$ such that $\phi'(u) = u'$, and moreover, $G_v(u) = T_{\phi(v)}(u')$. \qedhere
		\end{itemize}
	\end{lemma}
	\begin{proof}
		This follows from Lemmas \ref{lem:noInternalAliens}, \ref{lem:refinementLoop}, \ref{lem:refinementEdge}, \ref{lem:refinementNoExternalLeaf}, and \ref{lem:refinementAtMostOneAddedTree}.
	\end{proof}

	Now we are ready to prove that our algorithm will find a morphism of minimal degree. 
	\begin{theorem} \label{thm:existsPhiAlpha}
		Let $G$ be a graph with $\sgon(G) = k$. There is a tuple $\alpha$ such that $\phi_\alpha$ has degree $k$.
	\end{theorem}
\begin{proof}
	By Lemma \ref{lem:morphismProperties} we know that there is a refinement $H$ of $G$, a tree $T$ and a finite harmonic morphism $\phi\colon H\to T$ of degree $k$ which satisfy the properties in Lemma \ref{lem:morphismProperties}. By Section \ref{sec:boundingIndex}, we know that all indices are at most $\lfloor \frac{m-n+4}{2} \rfloor$.

	Define $T'$ as the subtree of $T$ that consists of the vertices $\phi(V(G))$.
	Notice that by Section \ref{sec:boundingSize}, $T'$ is connected and has at most $n$ vertices.
	Define $f\colon G \to T'$ as $\phi$ restricted to $V(G)$.
	Now let $r(uv)$ be the index that is assigned to every edge in $R_{uv}$.
	Let $\alpha$ be the tuple $(T', f,r)$.
	
	By the properties of $\phi$ and the construction of $\phi_\alpha$, it follows that $\phi_\alpha = \phi$. Thus $\phi_\alpha$ has degree $\sgon(G)$.  
\end{proof}

	\section{Stable gonality is in NP}\label{sec:NP}
	
	In this section we consider the decision problem ``\emph{stable gonality problem}'': 
	Given a graph $G$ and an integer $k$, does it hold that $\sgon(G) \leq k$? 
	\begin{theorem}
		The stable gonality problem belongs to the class NP. 
	\end{theorem}
	\begin{proof}
		Let $(G,k)$ be an instance of the stable gonality problem. 
		By Theorem \ref{thm:existsPhiAlpha} we know that there is a tuple $\alpha = (T,f,r)$ as in Section \ref{sec:algorithm} such that $\phi_\alpha$ has degree $\sgon(G)$.  
	This tuple has polynomial size. 
	Given a tuple $\alpha$, we can construct $\phi_\alpha$ in polynomial time, and we can compute its degree in polynomial time. 
	
	So for a yes-instance $(G,k)$, there is a tuple $\alpha = (T,f,r)$ with polynomial size and $\deg(\phi_\alpha) \leq k$, and we can check in polynomial time whether a tuple is a certificate for $(G,k)$. For a no-instance no such tuple exists. 
	\end{proof}

\section{NP-hard subproblem}

As described in Section \ref{sec:algorithm},
for a given graph $G$ our algorithm considers all tuples $\alpha = (T,f,r)$.
The tree $T'$ that is constructed from this tuple
and the vertices that will be internally added to $G$  do not depend on $r$.
The question arises whether
we can construct an $r$ such that $\phi_{(T,f,r)}$ has minimal degree
when we are already given the pair $(T,f)$.
This problem turns out to be NP-hard. 

For the proof we will use a reduction from the three-dimensional matching problem: Let $A, B, C$ be finite, disjoint sets, let $S\subseteq A\times B\times C$ and let $k$ be a natural number. Does there exist a set $M\subseteq S$ with $|M| \geq k$ such that every element of $A\cup B\cup C$ is contained in at most one tuple in $M$? We call such a set $M$ a \emph{matching}. 
This problem is known to be NP-hard, even when restricted to cases where $|A| = |B| = |C| = k$ and for every element of $A\cup B\cup C$ there are at least two tuple in $S$ containing this element \cite{Karp}. 

\begin{theorem} \label{thm:NPhardSubproblem}
	Given a graph $G$, a pair $(T,f)$ and an integer $k$. 
	The following problem is NP-hard: Does there exist a function $r:E(G) \to [\floor{(m-n+4)/2}]$ such that the morphism $\phi_{(T,f,r)}$ has degree at most $k$? 
\end{theorem}

\begin{proof}
	Let $(A_1,A_2,A_3,S,k)$ be an instance of the three-dimensional matching problem, where $|A_1| = |A_2| = |A_3| = k$
	and for every element of $A_1\cup A_2\cup A_3$ there are at least two tuples in $S$ containing this element. 
	Define $A = A_1 \cup A_2 \cup A_3$. 
	
	Construct the following graph $G$.
	Add two vertices $u_a$ and $v_a$ for every element $a\in A$ and add a vertex $w_s$ for every $s\in S$.
	Add an edge between $u_a$ and $v_a$ for every $a \in A$.
	For every tuple $s=\{x,y,z\}\in S$, we add edges $v_xw_s$, $v_yw_s$, $v_zw_s$ and edges $u_xv_x$, $u_yv_y$,
	and $u_zv_z$.
	See Figure \ref{fig:NPhardSubproblem} for an illustration. 
	\begin{figure}
	\centering
	\begin{tikzpicture}
	\node[vertex, label=below:$v_a$] (a) at (-1.9,.7) {};
	\node[vertex, label=right:$v_b$] (b) at (-1.5,1.3) {};
	\node[vertex, label=left:$v_c$] (c) at (1.5,1.3) {};
	\node[vertex, label=below:$v_d$] (d) at (1.9,0.7) {};
	\node[vertex, label=right:$v_e$] (e) at (0.3,-2) {};
	\node[vertex, label=left:$v_f$] (f) at (-0.3,-2) {};
	\node[vertex, label=left:$u_a$] (ua) at (-2.7,1.1) {};
	\node[vertex, label=$u_b$] (ub) at (-2.3,1.8) {};
	\node[vertex, label=$u_c$] (uc) at (2.3,1.8) {};
	\node[vertex, label=right:$u_d$] (ud) at (2.7,1.1) {};
	\node[vertex, label=right:$u_e$] (ue) at (0.3,-2.8) {};
	\node[vertex, label=left:$u_f$] (uf) at (-0.3,-2.8) {};
	\node[vertex, label=$w_q$] (w1) at (-.3,0.4) {};
	\node[vertex, label=$w_t$] (w4) at (.3,0.4) {};
	\node[vertex, label=right:$w_s$] (w3) at (0.3,-.3) {};
	\node[vertex, label=left:$w_r$] (w2) at (-0.3,-.3) {};
	\draw[edge] (w1) -- (a);
	\draw[edge] (w1) -- (c);
	\draw[edge] (w1) -- (e);
	\draw[edge] (w2) -- (a);
	\draw[edge] (w2) -- (c);
	\draw[edge] (w2) -- (f);
	\draw[edge] (w3) -- (b);
	\draw[edge] (w3) -- (d);
	\draw[edge] (w3) -- (e);
	\draw[edge] (w4) -- (b);
	\draw[edge] (w4) -- (d);
	\draw[edge] (w4) -- (f);
	\draw[edge] (ua) -- (a); 
	\draw[edge] (ua) to [relative,out=50, in=130] (a);
	\draw[edge] (a) to [relative,out=50, in=130] (ua);
	\draw[edge] (ub) -- (b); 
	\draw[edge] (ub) to [relative,out=50, in=130] (b);
	\draw[edge] (b) to [relative,out=50, in=130] (ub);
	\draw[edge] (uc) -- (c); 
	\draw[edge] (uc) to [relative,out=50, in=130] (c);
	\draw[edge] (c) to [relative,out=50, in=130] (uc);
	\draw[edge] (ud) -- (d); 
	\draw[edge] (ud) to [relative,out=50, in=130] (d);
	\draw[edge] (d) to [relative,out=50, in=130] (ud);
	\draw[edge] (ue) -- (e); 
	\draw[edge] (ue) to [relative,out=50, in=130] (e);
	\draw[edge] (e) to [relative,out=50, in=130] (ue);
	\draw[edge] (uf) -- (f); 
	\draw[edge] (uf) to [relative,out=50, in=130] (f);
	\draw[edge] (f) to [relative,out=50, in=130] (uf);
	\draw[->] (3.1,-0.5) -- (3.9,-0.5);
	\node[vertex, label=$v_1$] (v1) at (5.3,1) {};
	\node[vertex, label=$v_2$] (v2) at (8.7,1) {};
	\node[vertex, label=right:$v_3$] (v3) at (7,-2) {};
	\node[vertex, label=$u_1$] (u1) at (4.4,1.5) {};
	\node[vertex, label=$u_2$] (u2) at (9.6,1.5) {};
	\node[vertex, label=right:$u_3$] (u3) at (7,-2.8) {};
	\node[vertex, label=$w$] (w) at (7,0) {};
	\draw[edge] (w) -- (v1);
	\draw[edge] (w) -- (v2);
	\draw[edge] (w) -- (v3);
	\draw[edge] (u1) -- (v1); 
	\draw[edge] (u2) -- (v2); 
	\draw[edge] (u3) -- (v3); 
	\end{tikzpicture}
	\caption{An example of the graph $G$ and tree $T$ constructed in the proof of Theorem \ref{thm:NPhardSubproblem} for the sets $A_1 = \{a,b\}$, $A_2 = \{c,d\}$, $A_3 = \{e,f\}$ and $S = \{(a,c,e), (a,c,f), (b,d,e), (b,d,f)\}$.} \label{fig:NPhardSubproblem}
\end{figure}
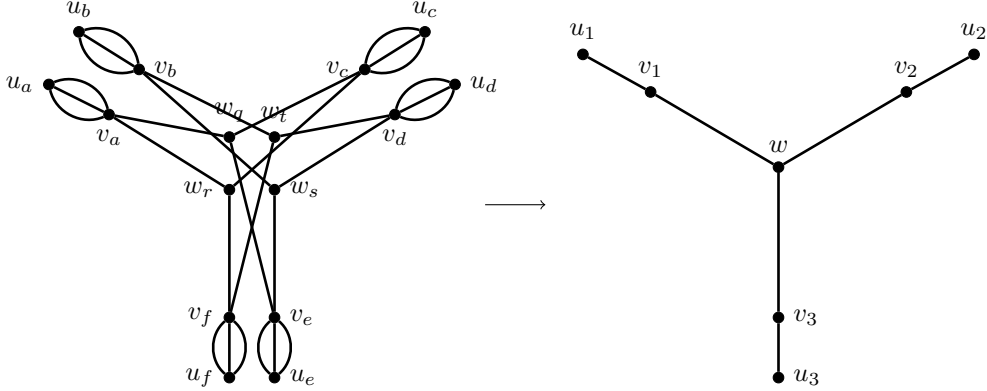 

	Let $T$ be the tree that consists of vertex $w$, vertices $u_{i}, v_{i}$ for $1\leq i\leq 3$, and edges $u_{i}v_{i}$ and $v_{i}w$ for $1\leq i\leq 3$.
	Define $f:V(G) \to V(T)$ as follows: 
	\begin{align*}
	f(x) = \begin{cases}
	w & \text{if $x=w_s$ for $s\in S$,} \\
	v_{i} & \text{if $x = v_a$ with $a\in A_i$,}\\
	u_{i} & \text{if $x = u_a$ with $a\in A_i$.}\\
	\end{cases}
	\end{align*} 

	Now $(G,T,f, |S|+k)$ is an instance of our problem. Notice that
	the degree of $\phi_{(T,f,r)}$ will be at least $|S|+k$ for any $r$,
	since at least $|S|+k$ edges map to each edge $u_{i}v_{i}$. 
	We will now prove that there is an $r$ such that $\phi_{(T,f,r)}$ has degree $|S|+k$ 
	if and only if there is a matching $M\subseteq S$ with $|M| \geq k$. 

	Given a perfect matching $M\subseteq S$, that is, a matching with $|M| = k$, we can find a $\phi$ of degree $|S|+k$
	by putting index $2$ on all edges $w_mv_a$ with $m\in M$ and $a\in m$.

	Now suppose that $\phi$ has degree $|S|+k$.
	This implies that each edge $u_av_a$ has index $1$ and that there are no externally added vertices mapped $u_{i}$ for all $i$. 
	Given a vertex $v_a$. Either one of the edges $v_aw_s$ has index $2$, or $v_a$ has an external added neighbour that is mapped to $w$. In the last case, there is an external added vertex mapped to one of the vertices $u_{i}$, which yields a contradiction. So, exactly one of the edges $v_aw_s$ must have index $2$; the others will have index $1$. 
	At each vertex $w_s$, all three edges will have the same index.
	This index must be either $1$ or $2$,
	since otherwise there will be an external added vertex that is mapped to a vertex $u_{i}$. 
	This implies that there are exactly $k$ vertices $s\in S$ with index $2$, while all other vertices $s$
	have index $1$. These $k$ tuples $(a_1, a_2, a_3)$ will form a matching, and hence we can solve the three-dimensional
	matching problem using the problem of finding an optimal $r$.
\end{proof}

	\section{Conclusion}
	
	Stable gonality is defined using three infinite loops: there are infinitely many refinements of a graph, there are infinitely many trees, and there are infinitely many finite harmonic morphisms from a refinement to a tree. In this paper we bounded the refinements, trees and morphism which we have to consider. This yields an algorithm to compute the stable gonality of a graph in $O\parens*{\poly(n,m) \cdot (1.33n)^n \parens*{\frac{m-n+4}2}^m}$ time. From these bounds and the algorithm it also follows that the stable gonality problem is in NP. 
	
	Some interesting questions remain open. Firstly: is there a faster algorithm to compute the stable gonality of a graph? Secondly, we do not know whether computing stable gonality is in XP or FPT, or whether it is W[1]-hard. Thirdly, are there problems which are untractable with treewidth as parameter, that are tractable with stable gonality as parameter? Lastly, is there a variant of Courcelle's theorem \cite[Chapter 13]{DowneyFellows} for graphs of bounded stable gonality?

\end{document}